# VQ-Wave: A physics-driven spatio-temporal deep learning approach for non-contrast-enhanced lung ventilation and perfusion MRI

Grzegorz Bauman[1,2], Pavlos Panos[1,2], Philipp Latzin[3], Oliver Bieri[1,2]

[1]Division of Radiological Physics, Department of Radiology, University of Basel Hospital, Basel, Switzerland

[2]Department of Biomedical Engineering, University of Basel, Basel, Switzerland

[3]Pediatric Respiratory Medicine, Department of Pediatrics, Inselspital, Bern University Hospital, University of Bern, Switzerland

**Corresponding author:**

Grzegorz Bauman, PhD
University Hospital Basel
Department of Radiology
Petersgraben 4
4031 Basel, Switzerland
E-mail: grzegorz.baumann@usb.ch
Phone: +41.61.556.57.28

**Running head:** Spatio-temporal DL for lung V/Q

# ABSTRACT

**Purpose:** To develop a robust deep learning framework for non-contrast-enhanced functional lung MRI, overcoming the limitations of spectral decomposition in the presence of physiological non-stationarity.

**Methods:** We introduce VQ-Wave (Ventilation/Q-perfusion Waveform-based Assessment of Variable Evolutions), a physics-driven spatio-temporal inception neural network trained on synthetic signal models to estimate ventilation and perfusion parameters. By processing local spatial context alongside temporal evolution, the network learns to robustly decouple physiological signals from noise. The training generator simulated realistic non-stationary dynamics, including amplitude modulations, frequency drifts, and noise. Performance was validated against matrix pencil (MP) decomposition using numerical phantoms and in-vivo functional lung MRI acquired in four healthy volunteers and two children with cystic fibrosis (CF) at 1.5T. Robustness was assessed across varying noise levels, physiological instabilities, and scan durations (truncating the acquisition from N=140 to N=40 images).

**Results:** In numerical benchmarks, VQ-Wave demonstrated superior robustness to non-stationarity, maintaining low global and regional error rates where MP exhibited stochastic instability due to spectral leakage. In-vivo, VQ-Wave accurately captured functional defects in patients with CF yielding diagnostically consistent ventilation and perfusion maps with high quantitative stability (global mean variation < 12%) even when scan time was reduced from 45s to 15s (N=40). Conversely, under irregular physiology and short scan lengths, MP decomposition severely degraded, exhibiting systematic amplitude instability, overestimation bias, and regional signal dropouts.

**Conclusion:** VQ-Wave offers a robust, physics-driven neural network-based alternative to spectral decomposition. By effectively handling physiological irregularity and noise, it enables reliable functional lung imaging with substantially shortened acquisition protocols.

## INTRODUCTION

In the recent years MRI has emerged as a promising, ionizing radiation-free modality for the assessment of pulmonary disease, offering an alternative to computed tomography (CT) or nuclear medicine modalities (1,2). While CT remains the gold standard for morphological lung imaging, MRI has established distinct clinical indications (3) - particularly for pediatric longitudinal monitoring and specific tissue characterization - yet the routine quantification of regional ventilation and perfusion remains largely dominated by nuclear medicine and contrast-enhanced modalities. Dynamic contrast-enhanced (DCE) MRI using gadolinium-based chelates remains the predominant MR method for perfusion imaging, favored for its robustness and straightforward implementation in clinical routine (4). However, its application in vulnerable populations is constrained by the invasiveness of venous access – a significant source of discomfort in pediatric care – and safety concerns regarding renal insufficiency, during pregnancy or long-term intracranial gadolinium deposition (5-7). Conversely, inhaled hyperpolarized noble gas MRI with $^{3}$He or $^{129}$Xe provides exquisite ventilation maps but faces significant barriers to widespread adoption due to high costs, logistical complexity, and the need for specialized hardware (8,9).

Consequently, significant research effort has been directed toward non-contrast-enhanced (proton-based) functional MRI techniques that exploit intrinsic signal modulations of the lung parenchyma caused by respiratory and cardiac cycles (10,11). One of the first approaches to leverage this principle was Fourier decomposition MRI, which applies voxel-wise spectral analysis to separate frequency components (12). This concept was further refined by matrix pencil (MP) MRI (13), employing linear prediction to improve frequency estimation, and Phase-Resolved Functional Lung (PREFUL) MRI (14), which utilizes retrospective phase sorting or SElf-gated Non-Contrast-Enhanced FUnctional Lung imaging (SENCEFUL) (15) technique employing non-phase-encoded direct current (DC) signal necessary for gating and spectral separation of physiological cycles. More recently, dynamic mode decomposition (DMD) has been proposed as a robust data-driven alternative for separation of the signal into spatial modes and their temporal evolution (16).

However, a fundamental limitation shared by the methods relying on spectral analysis and phase-sorting techniques is their reliance on the assumption of physiological stationarity or reproducible cyclic dynamics. These algorithms mathematically model respiratory and cardiac signals as stable oscillators or coherent modes. In clinical reality cardiopulmonary physiology is inherently non-stationary. Spontaneous deep breaths (sighs), respiratory rate drifts, and arrhythmias violate the steady-state assumption. In spectral and decomposition methods, such irregularities cause

"spectral leakage" or mode mixing, where signal energy is spread across multiple bins or modes, leading to amplitude underestimation or noise artifacts.

Deep learning (DL) offers a promising paradigm shift. Neural networks have already demonstrated efficacy in analyzing physiological time-series, such as extracting respiratory surrogates from k-space data or detecting cardiac phases (17-19). Unlike rigid mathematical models, neural networks can learn non-linear features and temporal dependencies from raw data. However, the application of DL to voxel-wise physiological signal modeling, specifically for the direct regression of ventilation and perfusion parameters from proton signal time-courses, remains underexplored. The primary bottleneck is the lack of ground truth, since it is impossible to measure the exact ventilation amplitude of a single voxel in a free-breathing patient to train a neural network. While synthetic lung models, have been introduced as valuable validation benchmarks, they are not designed as stochastic generators for the supervised learning of non-stationary dynamics (20).

In this work, we propose a novel spatio-temporal deep learning framework, termed VQ-Wave (Ventilation and Q-perfusion Waveform-based Assessment of Variable Evolutions), designed to robustly estimate lung ventilation and perfusion, specifically targeting the challenge of physiological non-stationarity. To overcome the lack of in-vivo ground truth, we introduce a synthetic physiological simulator capable of generating infinite variations of non-stationary breathing patterns including random amplitude modulation, frequency drifts, and noise to train the network. We hypothesize that a network trained on this stochastic "chaos" will learn to generalize to real-world in-vivo data, effectively functioning as a non-linear filter that outperforms standard spectral methods such as MP decomposition in terms of leakage suppression and stability. We validate this approach using a dynamic 2D (2D+t) numerical phantom and demonstrate its clinical feasibility in four healthy volunteers and two pediatric patients with cystic fibrosis.

## METHODS

### Signal Model

The temporal evolution of the MR signal intensity $y(t)$ in a magnitude voxel is modeled as the absolute value of a complex signal consisting of a static baseline, dynamic physiological components, and complex Gaussian noise:

$$y(t) = \left|B + S_v(t) + S_q(t) + \eta(t)\right| \quad \text{[Eq. 1]}$$

where $B$ – represents the static tissue baseline, $\eta$ – complex Gaussian noise, and $S_v(t)$ and $S_q(t)$ represent the ventilation and perfusion signal components, respectively.

To capture physiological irregularities such as deep breaths (sighs), heart rate variability, and respiratory rate drift, we define each dynamic component $c \in \{v, q\}$ as an amplitude- and frequency-modulated oscillator by using a modified Lujan motion model (21):

$$S_c(t) = p_c \cdot A_c \cdot a_c(t) \cdot \left(\frac{1+\cos(\Phi_c(t))}{2}\right)^{\gamma_c} \quad \text{[Eq. 2]}$$

where:

- $A_c$ is the fundamental amplitude (local change of parenchyma density or perfusion).
- $p_c$ is the polarity parameter: $p_v = -1$ for ventilation (representing inspiratory proton density reduction) and $p_q = +1$ for perfusion (representing inflow enhancement).
- $a_c(t)$ is the time-varying amplitude envelope, normalized to 1.0 for stationary breathing or perfusion, simulating transient events such as deep breaths or pulse pressure variations.
- $\Phi_c(t)$ is the instantaneous phase, the integral of the time-varying frequency $f_c(t)$ plus random initial phase offset $\phi_{0,c}$:

$$\Phi_c(t) = 2\pi \int_0^t f_c(\tau) d\tau + \phi_{0,c} \quad \text{[Eq. 3]}$$

- $\gamma_c$ is the Lujan shape exponent. For ventilation, we sample $\gamma_c \in \{1.0, 2.0, 4.0\}$; a higher exponent combined with negative polarity creates sharp inspiratory signal voids and prolonged expiratory plateaus. For perfusion, we typically set $\gamma_c \approx 1$ to model sinusoidal pulsatile flow.

Figure 1 provides a representative visualization of the stochastic modulation parameters.

### Synthetic Training Data

Estimating the parameters of the non-stationary signal model (Eq. 2) from noisy, short-time data series is an ill-posed inverse problem. To address this, we employ a supervised deep learning framework. However, obtaining ground-truth physiological labels from in-vivo scans is not feasible due to the lack of voxel-wise reference standards for non-contrast ventilation and perfusion MRI methods. To this end, we developed a synthetic data generator to create training samples with known ground truth. This framework can generate infinite variations of our signal model, ensuring the learning algorithm captures the underlying physiological coupling rather than memorizing specific anatomical features.

Training samples are generated as 3×3 voxel patches to provide local spatial context. For each patch, the central voxel's signal properties including baseline intensity, physiological modulation frequencies, and pathological defect probabilities are sampled from specific distributions representing distinct tissue classes and noise profiles (c.f. Supporting Information Table S1). The parameter probability distributions provided in the table S1 serve as data augmentation boundaries rather than epidemiological prevalences. To prevent class imbalance and bias toward healthy tissue, pathological voids and complex edge cases were deliberately over-sampled. Furthermore, continuous physiological variables were sampled from broad, uniform distributions. This strategy forces the network to learn the underlying MR signal physics rather than memorizing statistical priors, ensuring robust generalization across the entire physically possible parameter space.

To ensure that the network learns to distinguish anatomical boundaries while preserving structural sharpness, we arranged the selected tissue classes into three distinct spatial configurations:

- Homogeneous patches (~60%) where all 3×3 voxels belong to the same class. This teaches the network to utilize spatial averaging for denoising in consistent regions.
- Boundary patches (~30%) containing a sharp transition (step-function) between two different classes, such as the lung-diaphragm interface or the vessel-parenchyma interface). This configuration forces the network to function as a non-linear arbiter, learning to selectively gate neighbor information to prevent excessive smoothing across distinct tissue boundaries.
- Small structures (~10%) where the center voxel differs from the surrounding neighbors (e.g. a small vessel cross-section within the parenchyma), ensuring the network can detect small features without washing them out.

By training on these spatial topologies, the network learns to balance local denoising with spatial specificity.

### Neural Network Architecture

We propose a custom deep learning architecture, the spatio-temporal inception network (Figure 2), designed to estimate the physiological signal parameters ($A_v, A_q, f_v, f_q, \phi_v, \phi_q$) directly from the motion-corrected MR time-resolved ultra-fast bSSFP series. The network's input is constructed as a multi-channel tensor of shape $N_{ch} \times T$, where $T$ represents the number of time points. To encode local spatial context, we flatten the 3×3 spatial neighborhood into the channel dimension (preserving spatial topology through deterministic channel ordering). Thus, the input consists of 9 channels representing the normalized signal time-courses of the target voxel and its neighbors, plus an additional explicit channel encoding the imaging rate to ensure the network remains invariant to acquisition speed. We employed the feed-forward convolutional architecture rather than recurrent or transformer-based models (22,23) because the task of quantifying global physiological parameters from quasi-periodic signals is effectively a spectral estimation problem best solved by learnable non-linear filter banks (24), avoiding the computational overhead of sequential forecasting models.

The core feature extractor utilizes a stack of 1D inception blocks with varying kernel (25). This multi-scale design allows the network to simultaneously capture high-frequency physiological signals, such as the cardiac cycle, while large kernels integrate information over longer time windows to resolve low-frequency respiratory trends and drifts. To further enhance feature selectivity, each inception block is integrated with a squeeze-and-excitation (SE) module (26). The SE block adaptively recalibrates channel-wise feature responses by explicitly modeling interdependencies between channels. In our context, this mechanism acts as a learned attention gate, allowing the network to suppress non-informative channels (e.g., noisy neighbors or static tissue signals) and emphasize relevant physiological waveforms. The feature extraction stage terminates in a hybrid temporal aggregation layer designed to capture both time-invariant and instantaneous signal properties. To estimate amplitudes $(A_v, A_q)$ and frequencies $(f_v, f_q)$, which are independent of the starting time-point, we utilize global average pooling to condense the temporal dimension into a noise-robust feature vector. Conversely, since phase is inherently position-dependent, global averaging would destroy the necessary timing information. Therefore, we simultaneously extract the feature vector corresponding to the temporal center of the scan. These two feature representations (averaged and centered) are concatenated and fed into fully connected heads that output the estimated amplitudes, frequencies, and phase components $(\sin\phi_{0,v}, \cos\phi_{0,v}, \sin\phi_{0,q}, \cos\phi_{0,q})$ for both ventilation and perfusion compartments.

The network was implemented in PyTorch (27) and trained on $10^6$ synthetic samples using the AdamW optimizer. We optimized a composite loss function:

$$\mathcal{L}_{total} = \lambda_A \mathcal{L}_A + \lambda_f \mathcal{L}_f + \lambda_\phi \mathcal{L}_\phi \quad \text{[Eq. 4]}$$

designed to balance parameter accuracy with physical constraints. The amplitude term $\mathcal{L}_A$ utilizes a Smooth-L1 loss incorporating a spatial "ghost penalty" to enforce sparsity and suppress leakage artifacts. Conversely, the frequency $\mathcal{L}_f$ and phase $\mathcal{L}_\phi$ error terms are signal-weighted by the ground-truth amplitude; this ensures the network prioritizes estimation in regions with valid physiological signals while ignoring the unstable phase of the background noise. The model was trained with empirically determined balancing coefficients $\lambda_A = 100$, $\lambda_f = 10$, $\lambda_\phi = 10$ to prioritize accurate amplitude quantification of the mean physiological state while maintaining spectral fidelity. A summary of the network optimization and training hyperparameters is provided in Supporting Information Table S2. The PyTorch code defining the network topology, pre-trained model weights and inference framework are available for non-commercial research purposes at: https://github.com/randomehro/vq_wave.

### Lung Phantom

To quantitatively validate the network, we developed a 2D+t numerical lung phantom framework $(256 \times 256 \times T)$ simulating spatially coherent ventilation and perfusion signal dynamics using the modulated oscillator model (Eq. 2). All parenchymal voxels share a synchronized global frequency trajectory $(f_v, f_q)$ and amplitude modulation envelope, while preserving local phase differences. To avoid data leakage, the physiological parameters assigned to the phantom were independently sampled from continuous but overlapping distributions. The phantom geometry is defined by a lung mask including parenchymal tissue and major pulmonary vessels derived from an automatically segmented in-vivo ultra-fast balanced steady-state free precession scan using neural network (28). Within these anatomical masks, we assigned ground-truth physiological parameters to model tissue heterogeneity. The parenchyma was modeled with a radial ventilation gradient $(A_v)$ decaying from the lung center to the periphery, while the vascular mask was assigned high baseline intensity and strong perfusion signal $(A_q)$ with zero ventilation $(A_v = 0)$.

We generated three benchmarks to test the model against matrix pencil decomposition: (1) noise robustness (varying noise levels), (2) physiological non-stationarity (varying amplitude modulation index α and frequency variability) and (3) temporal sensitivity (varying scan duration).

### In-vivo MRI scans

To demonstrate clinical feasibility, free-breathing lung functional MRI scans were performed on a commercial whole-body 1.5T MR-scanner (MAGNETOM Avanto-Fit and Sola-Fit, Siemens Healthineers, Forchheim, Germany) using 12-channel thorax and 24-channel spine coil as receiver and body coil as transmitter. The study was approved by the Institutional Review Board, and written informed consent was obtained from each subject. Four healthy volunteers (two females, two males; age range: 26-43 years) and two children with cystic fibrosis (CF) (12-year-old female and 14-year-old male) were scanned during free-breathing with a time-resolved 2D ultra-fast balanced steady-state free precession (uf-bSSFP) sequence (29). Main sequence parameters were as follows: TE/TR = 0.68/1.55 ms, TA per image = 118 ms, TA per slice = 45 s, field-of-view = 450×450 mm$^2$, in-plane resolution = 3.5×3.5 mm$^2$, slice thickness = 12 mm, flip angle = 60°, bandwidth = 2056 Hz/px, 150 coronal images per slice, interval between images =

182 ms, parallel imaging GRAPPA factor 2. First 10 images were discarded from further analysis since they were acquired in the transient magnetization state. The respiratory motion was corrected using an elastic image registration (30) and subsequently the lung tissue was segmented using an artificial neural network (28). The inference of the VQ-Wave model was directly implemented into the automated processing pipeline TrueLung (31) using the C++ torch library (27).

**Data evaluation**

Performance of the presented deep learning technique was validated using both the numerical phantom benchmarks and the in-vivo data. For the phantom experiments, we quantified the accuracy of the estimated ventilation and perfusion maps using the root mean squared error (RMSE) to measure the absolute deviation of the estimated parameters from the ground truth.

We performed a comparative analysis against MP decomposition across all three benchmarks (noise robustness, physiological non-stationarity, and temporal length sensitivity). MP decomposition was selected as a rigorous chronological time-domain baseline; alternative spectral methods (such as Fourier or dynamic mode decomposition) suffer equivalently from spectral leakage under non-stationary conditions, while retrospective phase-sorting techniques inherently enforce periodicity and smooth out transient irregular events. The MP reconstruction followed an optimized global workflow described previously (31) rather than a noise-susceptible voxel-wise approach. In this workflow, global respiratory and cardiac frequencies are first determined from the spatially averaged lung signal, followed by a voxel-wise least-square fit to resolve local amplitudes. In contrast, the proposed VQ-Wave network estimates physiological parameters purely from the local spatio-temporal context without global frequency priors. Furthermore, spatial fidelity was evaluated by calculating voxel-wise error maps (difference between the reconstructed maps minus ground truth) and linear regression scatter plots across the entire phantom domain under baseline, non-stationary, and truncated acquisition conditions.

For the in-vivo data, we evaluated the method's performance on datasets from both healthy volunteers and patients with CF. To evaluate performance under accelerated conditions, we reconstructed two distinct datasets for each subject: a reference standard utilizing the steady-state portion (N=140 images) of the full 45s acquisition, and a truncated initial subset using the steady-state portion (N=40 images) of a 15s acquisition. We performed a qualitative assessment of the reconstructed ventilation and perfusion maps to identify artifacts or signal dropout in the short acquisition. To further verify the robustness of the scan-time reduction, we performed a quantitative stability analysis by calculating the global mean signal amplitude across the entire lung mask for varying scan lengths incrementally truncated from N=140 down to N=40. The standard deviation of these global mean measurements was compared between VQ-Wave and the MP reference to quantify the stability of the physiological estimation with respect to the observation window length.

To generate quantitative functional maps, the raw amplitudes estimated by both methods were linearly scaled using established physiological models. Fractional ventilation (FV) was calculated as:

$$FV=\frac{A_v}{B+A_v/2-BG}\cdot 100\ [\%] \qquad \text{[Eq. 5]}$$

where $A_v$ is the estimated ventilation amplitude, $B$ is the baseline signal intensity (DC component), and $BG$ is the background noise. The quantitative pulmonary perfusion $Q$ was estimated using the method proposed by Kjørstad et al. (32) and expressed as:

$$Q=6000\cdot\frac{\overline{f_q}}{2}\ \frac{A_q}{A_{blood}}\left[\frac{\text{ml}}{\text{min}\cdot 100\text{ml}}\right] \qquad \text{[Eq. 6]}$$

where $A_q$ is the estimated perfusion amplitude, $\overline{f_q}$ is the global averaged cardiac frequency, and $A_{blood}$ is the reference amplitude measured in an artery representing voxels 100% filled with blood.

## RESULTS

Validation of VQ-Wave against the global matrix pencil reference using independent, simulated single-pixel time-courses with stationary properties (noise level of σ=5) is shown in Figure 3. High amplitude linearity (r ≥ 0.997) is observed for both ventilation and perfusion. A minor positive rectification bias is present in perfusion amplitude analysis for VQ-Wave near the noise floor, a consequence of estimating non-negative magnitudes without reliance of global frequency priors. For this reason, Bland-Altman analysis also reveals a negligible bias for VQ-Wave. Regarding phase recovery (Figure 3E-F), VQ-Wave leverages spatiotemporal context to achieve precise phase locking. The distribution of absolute phase errors follows a tight, half-normal profile with a RMSE of 3.3° for ventilation and 8.3° for perfusion. In contrast, MP suffers from significant phase randomization due to local noise dominance, yielding a broad, heavy-tailed distribution RMSE ≈70° where estimation errors approach uniformity.

Reconstruction performance on a spatially coherent anatomical lung phantom under high noise conditions (σ=11) is presented in Figure 4. For the high-amplitude ventilation maps (top row), both VQ-Wave and the MP reference successfully recover the underlying respiratory structure.

However, VQ-Wave achieves a higher visual signal-to-noise ratio, resulting in a more homogeneous amplitude distribution across the parenchyma compared to the grainier appearance of the MP map. The performance gap is more visible in the low-SNR perfusion regime: while MP exhibits characteristic noise amplification and graininess in the parenchyma, VQ-Wave effectively suppresses this noise to yield a coherent vascular map. Zoomed insets highlight the edge preservation capabilities of the network. In the ventilation map, the pulmonary vessels appear as sharp and dark signal voids, while in the perfusion map as high-signal structures with no evidence of the "halo" artifacts or blurring that typically result from standard 3×3 convolution filters. This confirms that the network utilizes spatial context for non-linear feature extraction rather than simple averaging.

Quantitative benchmarking of the proposed architecture against the MP reference across three axes of degradation is summarized in Figure 5. In the noise sweep, VQ-Wave consistently outperforms MP in the clinically relevant regime, reducing RMSE at noise levels ($\sigma=7$ to $\sigma=13$). While MP approaches VQ-Wave's performance at extreme noise levels ($\sigma>15$), this regime represents non-diagnostic quality where both methods degrade. Results from the instability benchmark, designed to model realistic respiratory variations with persistent background frequency drift and varying indices of amplitude modulation (AM), show a high baseline error for MP even at zero AM. This confirms the sensitivity of MP to non-stationary carriers. Furthermore, as modulation depth increases, MP displays erratic variance oscillating between moderate and severe error states. This stochastic instability arises because AM introduces spectral sidebands that the global spectral decomposition misinterprets as competing poles. Conversely, VQ-Wave demonstrates a near-flat error profile, indicating robustness to both frequency drifts and envelope modulations. Finally, the scan length benchmark highlights the efficiency of the deep learning approach. VQ-Wave converges to a stable solution with as few as 40 frames, whereas MP requires significantly longer series (>80 frames) to achieve comparable stability. This suggests that VQ-Wave can enable substantially shorter free-breathing scan protocols.

Spatial fidelity analysis performed on the numerical lung phantom for the estimation of ventilation and perfusion is presented in Figures 6 and 7. The spatial error maps indicate that the conventional MP estimator exhibits localized inaccuracies when processing non-stationary signals; specifically, contiguous zones of signal underestimation are observed during frequency drift (middle row), while scan truncation (N=40) results in noise amplification and spatial errors (bottom row). In contrast, VQ-Wave error maps show uniform accuracy across these scenarios. The voxel-wise scatter plots further characterize the fidelity and edge preservation of the two methods. While MP exhibits boundary leakage and "halo" artifacts around high-signal structures, associated with reduced regression slopes ($m < 1.0$), VQ-Wave maintains a near-perfect identity slope and negligible boundary leakage, indicating the resolution of sharp boundaries without spatial blurring.

Initial in-vivo feasibility was confirmed in healthy subjects. Supporting Information Figure S1 shows exemplary fractional ventilation and perfusion images acquired in a volunteer. While both VQ-Wave and MP yielded consistent, diagnostic-quality functional maps under stable, long-acquisition conditions (N=140), truncating the acquisition to 15 seconds (N=40) caused significant noise amplification and signal dropout in the MP perfusion maps, whereas VQ-Wave maintained robust image quality. This performance gap widened further in the presence of the non-stationary physiology characteristic in a 14-year-old male patient with cystic fibrosis (CF) (Figure 8).

For the patient with CF, reducing the scan length to N=40 resulted in a severe visual degradation of the MP perfusion map, characterized by noise amplification and loss of vascular detail, while VQ-Wave successfully preserved structural coherence. The spectrogram reveals frequency drifts of both respiratory and cardiac cycles. Furthermore, the breathing curve shows strong respiratory amplitude modulation. Quantitative stability analysis via a scan length sweep demonstrates that VQ-Wave yields highly stable global mean ventilation and perfusion estimates regardless of scan duration, maintaining a variation of less than 12%. In contrast, the MP reference exhibits severe quantitative volatility, with variations reaching up to 60%.

The temporal reproducibility of the deep learning approach is further demonstrated in Figure 9, where reconstructions from truncated windows (N=40) taken from the beginning, middle, and end of the time-series remain visually and quantitatively consistent. Conversely, the MP reconstructions exhibit window-dependent artifacts and signal dropouts, confirming that spectral methods struggle to isolate weaker cardiac signals without a long, stable observation window.

Finally, Figure 10 presents a comparison on a CF dataset (N=140) acquired in for a 12-year-old female patient with CF, which exhibits stable physiological cycles with minimal frequency drift and amplitude modulation. Under these conditions, both methods successfully identify the exact same complex peripheral functional impairment. The resulting functional defect maps show strong morphological agreement, yielding comparable outcomes (VQ-Wave: VDP = 16.9%, QDP = 15.3%; MP: VDP = 16.5% and VDP = 17.9%). Comparison of the defect maps yielded a spatial overlap (Dice coefficient) of 0.96 for VDP and 0.85 for QDP.

## DISCUSSION

In this work, we introduced a physics-driven deep learning framework, termed VQ-Wave, for the robust estimation of pulmonary ventilation and perfusion signal amplitudes from non-contrast-enhanced proton MRI. Unlike common supervised learning strategies that rely on image-based training, our network was trained entirely on synthetic signal models, allowing it to learn the underlying physical coupling of cardiopulmonary signals independent of scanner-specific image contrast or resolution. The proposed spatio-temporal architecture demonstrates superior robustness against noise and physiological non-stationarity compared to the matrix pencil decomposition, enabling reproducible functional imaging even under shortened acquisition protocols.

The primary advantage of the proposed VQ-Wave framework over established spectral methods, such as Fourier decomposition or matrix pencil, lies in its intrinsic robustness against physiological non-stationarity. Conventional spectral algorithms inherently model the respiratory and cardiac cycles as stable, periodic oscillators. As shown in our simulations, when this assumption is violated (e.g., due to respiratory rate drifts, arrhythmias, or spontaneous deep breaths), these methods suffer from spectral leakage, where signal energy is spread across multiple frequency bins. A related consideration applies to phase-resolved techniques (e.g. PREFUL or SENCEFUL). These methods retrospectively sort the entire time-series to reconstruct and often harmonically fit a single, representative physiological cycle. While highly effective for stable physiology, this binning process is generally optimized for conditions where cycle-to-cycle amplitude variations are minor and most beats or breaths contribute to reproducible mean state. Consequently, pronounced transient or irregular events such as the respiratory amplitude variations or cardiac frequency drifts observed in our CF patient data, might be smoothed out or treated as outliers during the sorting process. In contrast, VQ-Wave operates as a non-linear, time-domain filter that processes the signal evolution sequentially. By tracking local morphological features without requiring a strictly periodic model, the network demonstrates an improved capacity to preserve amplitude-modulated signals and mitigate the effects of non-stationary dynamics, thereby reducing the reliance on temporal averaging.

Furthermore, spatial fidelity analysis on the numerical lung phantom confirms that the global stability of VQ-Wave is not a result of compensatory regional masking. Voxel-wise error mapping demonstrates consistently low regional errors across the simulated field of view, ensuring that localized over- and under-estimations do not artificially stabilize the global mean. The network effectively prevents boundary leakage and preserves sharp physiological transitions, such as the vessel-parenchyma interface, without relying on spatial blurring, even under severe signal modulation and frequency drift.

This stability has direct implications for clinical workflow efficiency and diagnostic reliability in patients. Current spectral protocols typically require relatively long acquisition times (~140- 200 images per slice) to ensure sufficient spectral resolution and to average out background noise. Our in-vivo validation in pediatric patients with CF explicitly illustrates the vulnerability of these conventional methods to shortened scan times. When subjected to irregular cardiopulmonary dynamics of a patient with CF, truncating MP reconstruction to 40 frames triggered a failure mode, yielding noise-dominated maps and artifactual amplitude fluctuations. In contrast, VQ-Wave yielded markedly improved global mean values and preserved structural coherence for both ventilation and perfusion within the same acquisition window.

This capability suggests that the total acquisition time per slice could be reduced by a factor of 3, minimizing bulk motion artifacts and improving compliance in pediatric or dyspneic patients. Furthermore, the robustness of the method addresses a key challenge in longitudinal studies: tracking absolute functional decline over time. Importantly, testing on a high-quality CF dataset confirmed that VQ-Wave accurately detects and preserves the exact same peripheral pathological voids as the full-length MP reference, demonstrating very good morphological agreement in diagnostic defect maps (VDP/QDP). Beyond these threshold-based spatial metrics, the substantial reduction in amplitude variability achieved by VQ-Wave may provide a more reliable absolute baseline for quantitative longitudinal assessments.

A unique feature of this framework is its reliance on synthetic training data. Because the network learns from a physics-based simulator rather than site-specific image data, it is theoretically agnostic to field strength and pulse sequence, provided the temporal resolution is sufficient to resolve the cardiac cycle. In this study, we demonstrated the method at 1.5T, but the absence of image-based domain adaptation implies it could be readily applied to 0.55T or 3T platforms without retraining, provided the input signals are normalized as described.

We acknowledge limitations in this study. While our initial clinical translation in pediatric patients with CF successfully demonstrated the recovery of pathological defects and resilience to non-stationary physiology, this remains a pilot in-vivo validation. The complex signal characteristics associated with a broader spectrum of severe pulmonary diseases may present irregular features not fully captured by the current synthetic simulator. Furthermore, the synthetic generator simplifies certain physical and geometric nuisances inherent to 2D free-breathing MRI, such as residual registration errors, through-plane motion, and complex partial volume effects. While intrinsic sequence physics (e.g., ultra-short bSSFP) and the network's localized spatial processing naturally mitigate off-resonance effects and coil sensitivity drifts, unmodeled geometric confounders remain a fundamental limitation of all 2D non-contrast functional MRI techniques. As with any data-driven reconstruction, there is a theoretical risk that the network could produce texture-based hallucinations in the absence of valid physiological signal; therefore, the diagnostic accuracy and generalizability of this framework must be rigorously evaluated in future large-cohort clinical trials against established external reference standards.

## CONCLUSION

VQ-Wave represents a robust alternative to spectral and phase-resolved decomposition methods for free-breathing functional lung MRI. By decoupling estimation accuracy from the strict requirement of physiological stationarity, the method enables rapid, high-quality ventilation and perfusion mapping. While initial clinical feasibility was successfully demonstrated in pediatric patients with cystic fibrosis, future work will focus on validating the diagnostic accuracy and longitudinal reproducibility of this technique in larger patient cohorts against established gold-standard modalities.

# FIGURES

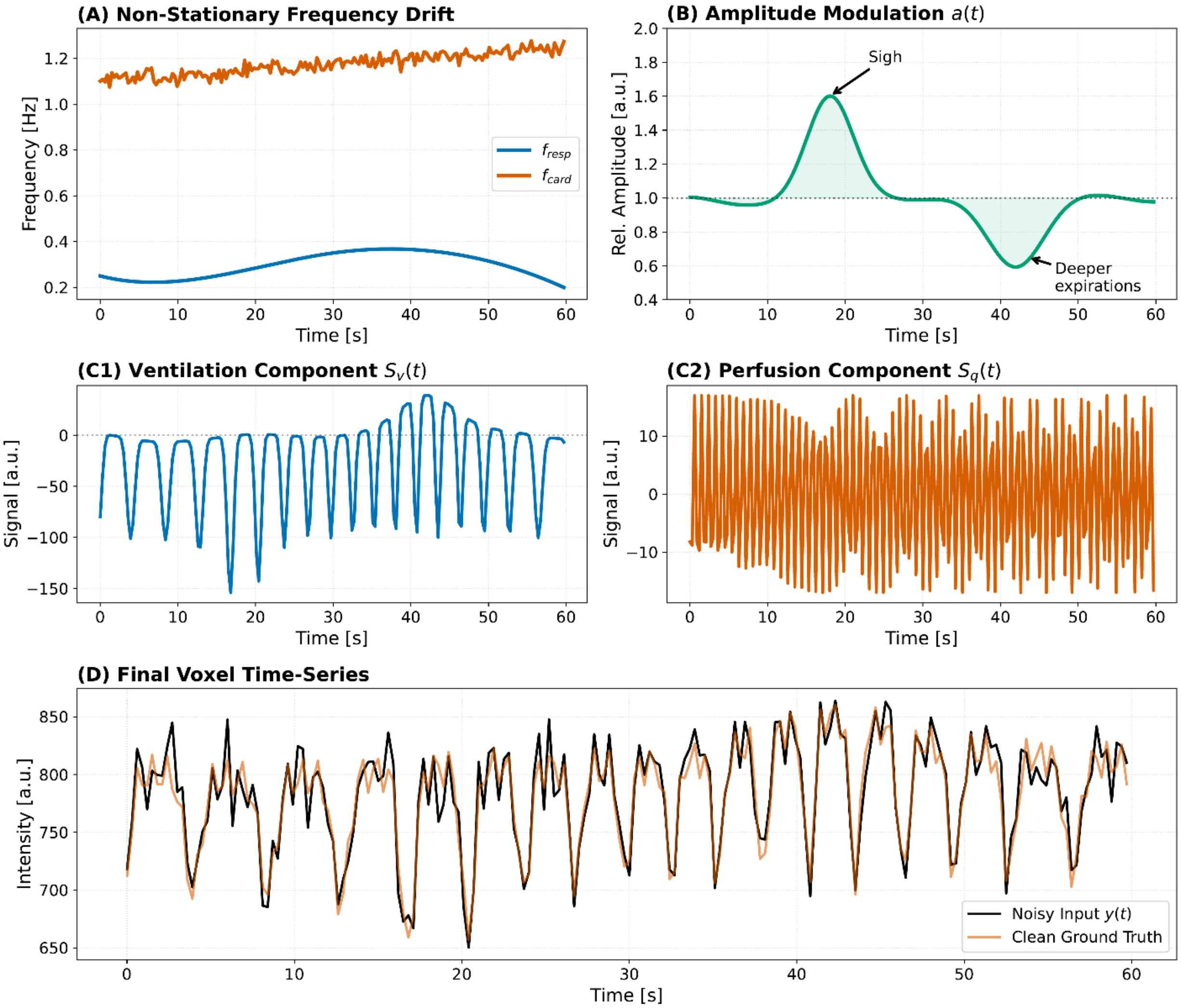

**Figure 1:** Generation of non-stationary physiological signal time-courses using the synthetic physiological simulator. (A) Evolution of instantaneous frequencies for ventilation ($f_v$, blue) and perfusion ($f_q$, red). Unlike standard spectral models that assume constant periodicity, the simulator introduces stochastic frequency drifts and heart rate variability to mimic natural physiological instability. (B) The respiratory amplitude modulation envelope $\alpha_v(t)$. This parameter simulates variable tidal volumes, including spontaneous deep breaths or deeper expirations. (C1-C2) Resulting signal components derived from the phase evolution. The ventilation waveform (C1) undergoes non-linear shaping modified Lujan model, $\gamma_v = 2$ to reproduce the temporal asymmetry of respiration. The perfusion component (C2) follows a pulsatile flow pattern. (D) The final synthetic MRI signal $y(t)$ (black line) generated by superimposing the components onto a baseline with added noise with the overlaid ground truth (orange line).

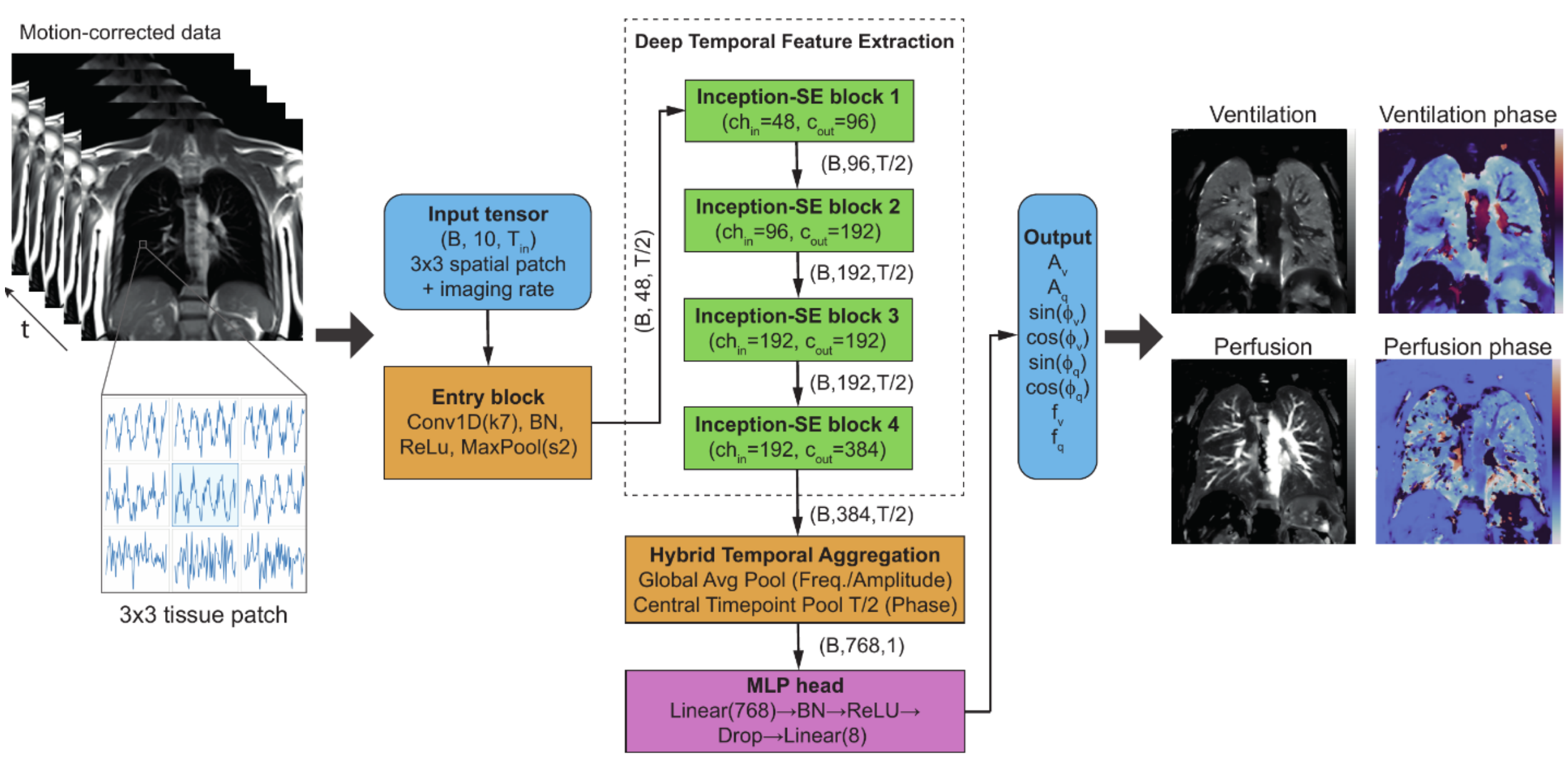

**Figure 2.** Schematic overview of the network architecture. The network processes inputs consisting of dynamic $3 \times 3$ spatial neighborhoods flattened into 9 channels, plus an additional channel for imaging rate, over a fixed temporal window of 40 to 190 timepoints ($T_{in}$). An initial convolutional stem layer reduces temporal dimensionality. The feature extraction backbone consists of four stacked inception squeeze-and-excitation (Inception-SE) blocks with progressively increasing channel widths. Input features are processed in parallel by four convolutional branches with varying temporal kernel sizes ($k = \{3,11,21,41\}$) to capture multi-scale dynamics. The outputs are concatenated, subjected to channel mixing via a $1 \times 1$ convolution, and adaptively recalibrated by a SE attention mechanism before being added to the residual skip connection. Finally, a regression head utilizes a hybrid pooling strategy combining global average pooling (for amplitude/frequency stability) and central time-point (to preserve instantaneous phase information) followed by a series of fully connected layers with dropout to predict the eight target parameters.

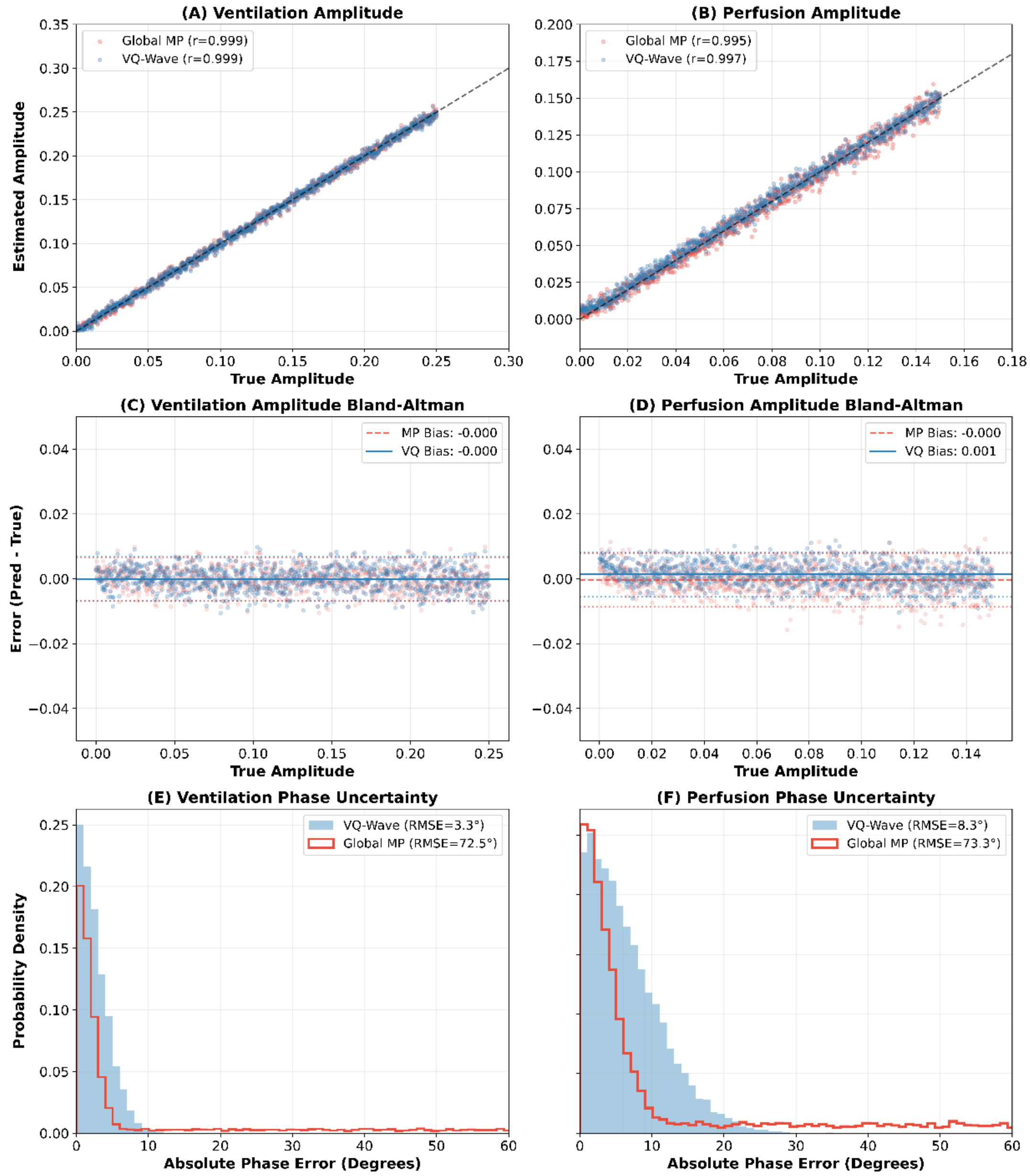

**Figure 3:** Quantitative validation of physiological parameter recovery. Performance of the proposed VQ-Wave network (blue) is compared to the reference global matrix pencil (MP, red) method using stationary physiological signal simulations at a uniform noise level of σ=5.0. (A, B) Scatter plots demonstrating the linearity of recovered amplitude for ventilation and perfusion channels. Pearson correlation coefficients (r) are provided to quantify agreement with the ground truth. (C, D) Bland-Altman analyses characterizing the amplitude estimation error. The solid lines denote the mean measurement bias, while the dashed lines represent the 95% limits of agreement (±1.96 SD). (E, F) Histograms of absolute phase error. VQ-Wave exhibits a narrow, half-normal error distribution, whereas MP displays a heavy-tailed distribution indicative of phase instability. Legend values denote the root mean square error (RMSE)

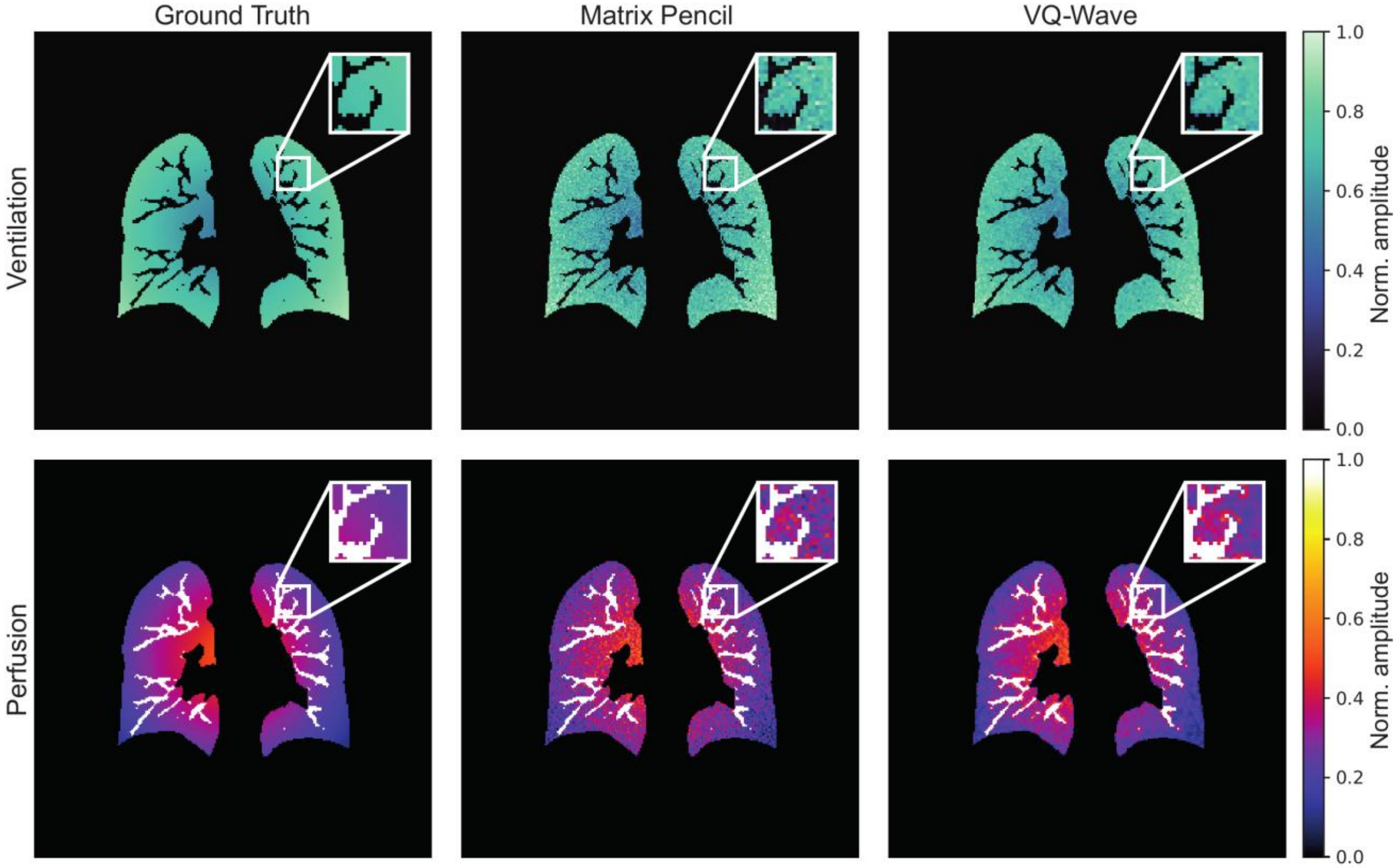

**Figure 4:** Qualitative assessment of spatial reconstruction fidelity on the numerical lung phantom. Representative amplitude maps generated from ground truth signals under high noise conditions ($\sigma$=11.0). Columns compare the (left) ground truth reference, (center) matrix pencil reconstruction, and (right) the proposed VQ-Wave network. Zoomed insets focus on fine anatomical structures of pulmonary vessels. While both methods attempt to recover the high-amplitude ventilation structure (top row), VQ-Wave achieves significantly higher SNR through superior denoising. In the low-amplitude perfusion maps (bottom row), the MP reconstruction suffers from marked graininess and noise-floor artifacts, whereas VQ-Wave demonstrates robust signal recovery. Crucially, VQ-Wave preserves sharp transitions at vessel boundaries—visible as sharp signal voids in ventilation and bright structures in perfusion—confirming that the network's 3×3 spatial integration does not induce blurring artifacts even at high noise levels.

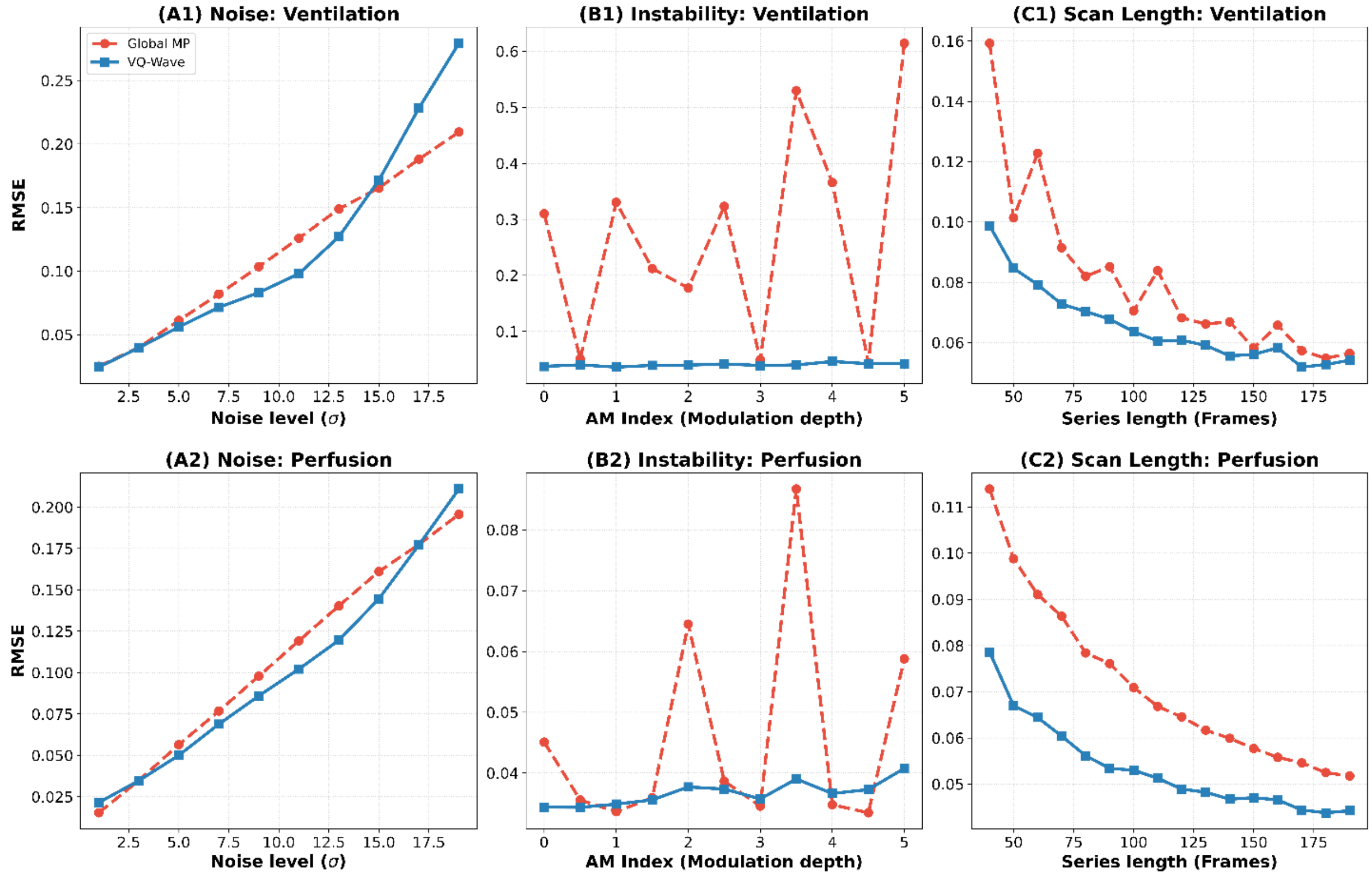

**Figure 5.** Benchmarking of reconstruction robustness. Comparison of root mean squared error (RMSE) for VQ-Wave (blue squares) and matrix pencil (MP, red circles) across three perturbation regimes. (A1-A2) Noise sensitivity: VQ-Wave maintains lower reconstruction error across a wide range of noise levels (σ=1 to 15), demonstrating superior denoising capabilities compared to MP. (B1-B2) Sensitivity to physiological instability: Signals were simulated with a constant background frequency drift (~15% of the frequency deviation for both respiratory and cardiac cycles) combined with varying amplitude modulation (AM) indices. The high baseline error of MP at AM=0 reflects its inability to track even this moderate frequency drift. As AM increases, MP exhibits erratic error spikes ("stochastic instability") caused by interference from AM-induced spectral sidebands. In contrast, VQ-Wave remains invariant to both frequency drift and envelope modulation. (C1-C2) Scan efficiency: VQ-Wave converges to a low-error solution with short acquisition series of $N \approx 40$ (frames). In contrast, MP requires significantly longer observation windows to stabilize the spectral covariance matrix.

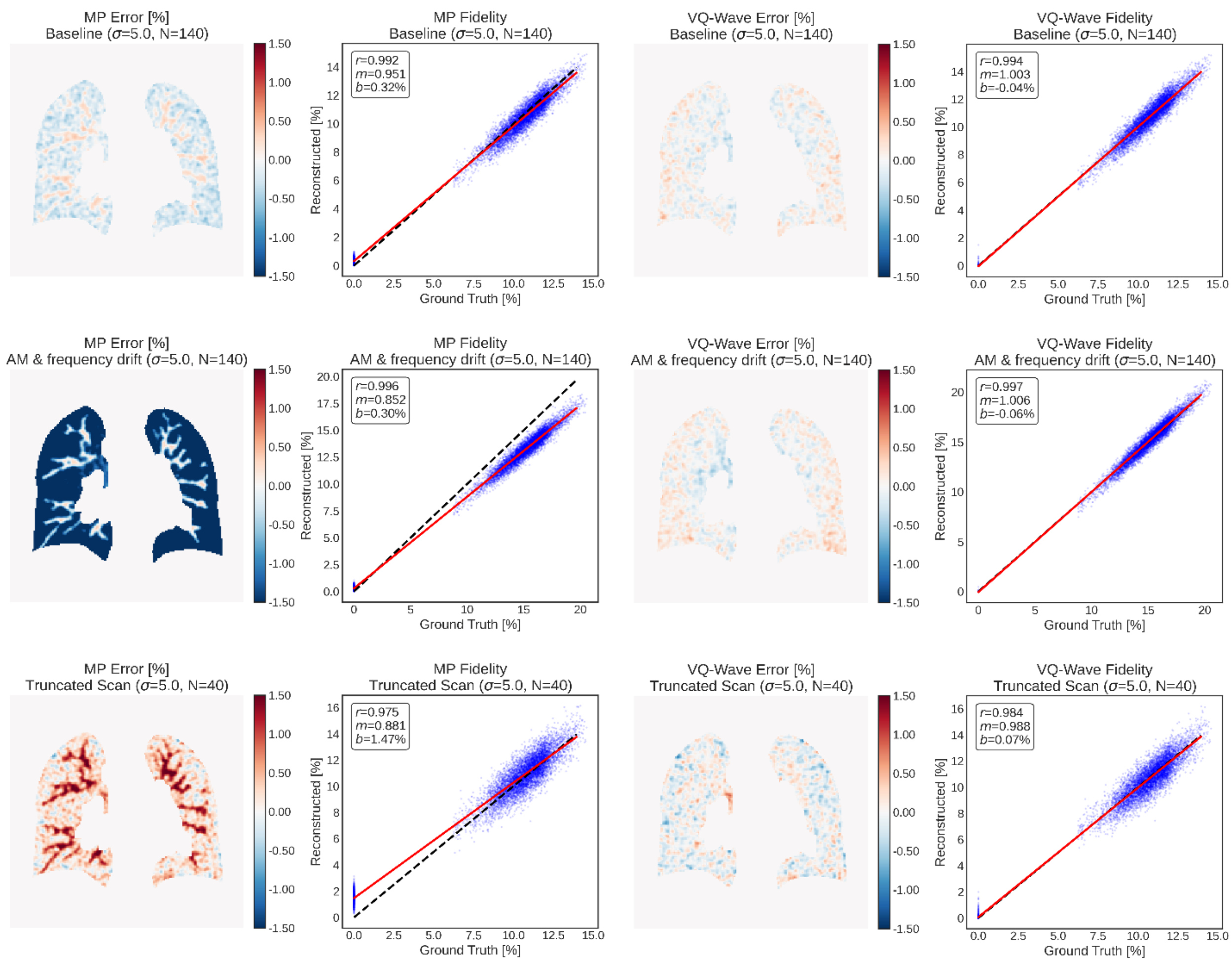

**Figure 6.** Spatial fidelity of ventilation reconstruction. Comparison of reconstruction accuracy between the conventional matrix pencil algorithm (MP, left block) and the proposed VQ-Wave network (right block) against numerical ground truth. Spatial error maps (reconstructed minus ground truth amplitude) and voxel-wise scatter plots are evaluated under three simulated scenarios: an ideal baseline steady-state acquisition (scan length N=140, top row), high non-stationary physiological drift with erratic amplitude modulation (N=140, middle row) and scan shortening (N=40, bottom row). Scatter plots include the ideal identity line (dashed black) and linear regression fit (solid red), with corresponding Pearson correlation r, slope m, and intercept b.

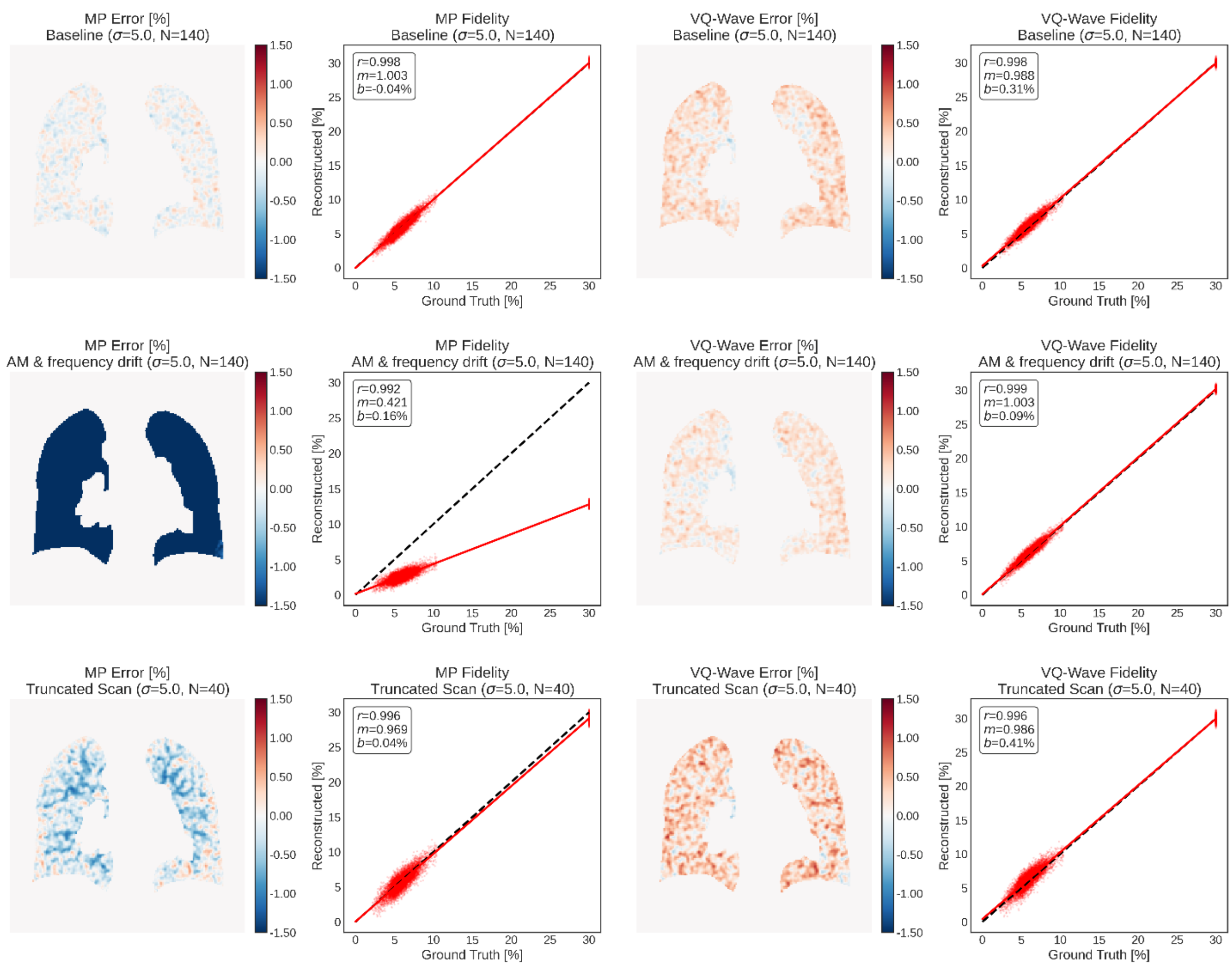

**Figure 7.** Spatial fidelity of perfusion reconstruction. Evaluation of perfusion quantification accuracy comparing the matrix pencil (MP) reference and VQ-Wave. Similar to the ventilation accuracy quantification, spatial error maps and voxel-wise correlation plots are provided for baseline (scan length N=140, top row), non-stationary simulated regimes (N=140, middle row) and truncated scan (N=40, bottom row). Scatter plots include the ideal identity line (dashed black) and linear regression fit (solid red), with corresponding Pearson correlation r, slope m, and intercept b.

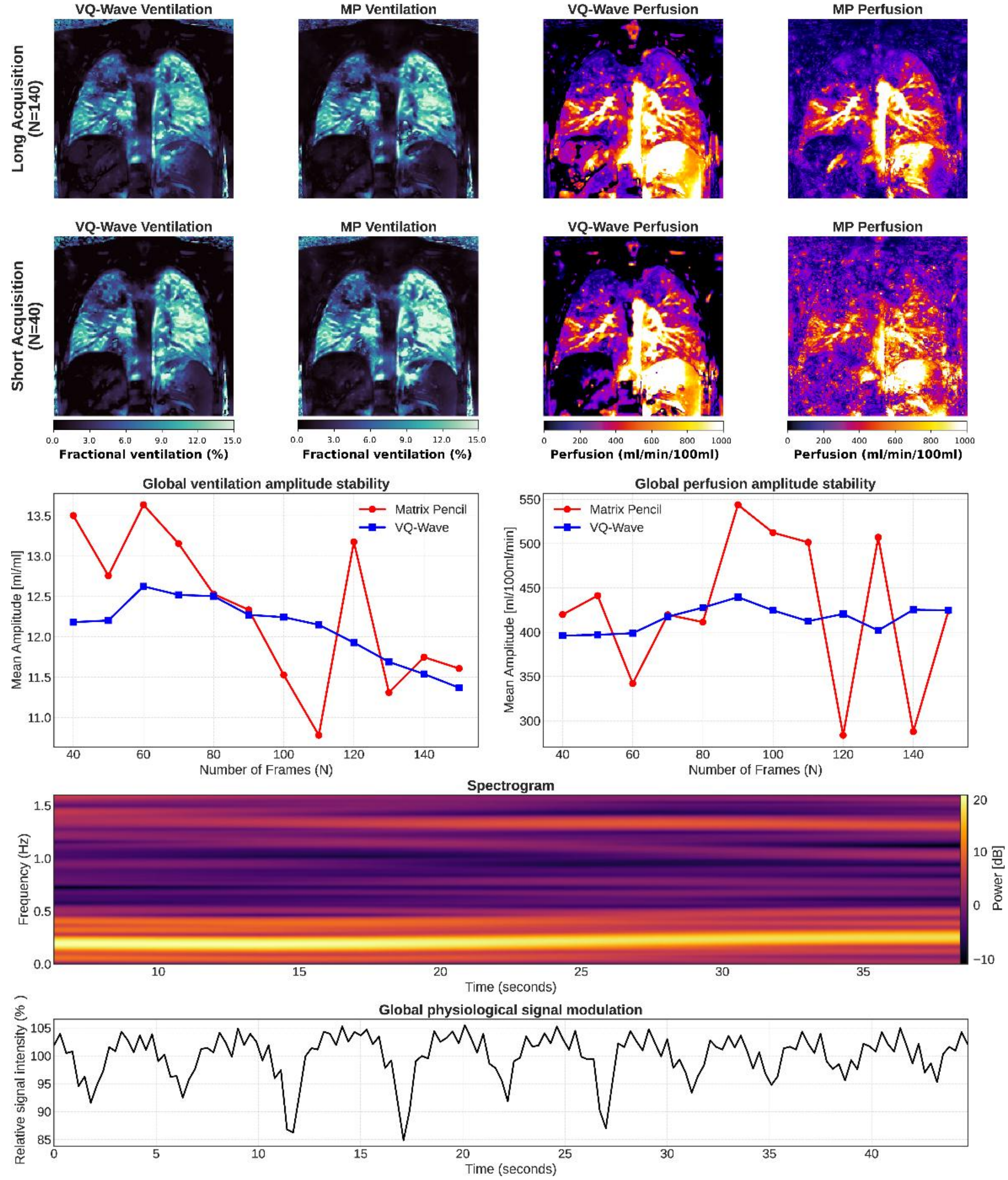

**Figure 8.** In-vivo performance and quantitative stability under physiological non-stationarity. Evaluation of a 14-year-old male with cystic fibrosis exhibiting irregular cardiopulmonary dynamics. Clinical lung function tests indicated a forced vital capacity (FVC) of 100%, forced expiratory volume in 1 second (FEV1) of 85%, and a lung clearance index at 2.5% (LCI 2.5%) of 13.02. (Top rows) Qualitative comparison of fractional ventilation and perfusion maps generated by VQ-Wave and the matrix pencil (MP) reference for a full-length baseline acquisition (N=140) versus a truncated acquisition (N=40). (Middle row) Quantitative stability analysis depicting the global mean ventilation and perfusion amplitudes as a function of the reconstructed scan length (from N=40 to N=140). (Bottom rows) Time-frequency spectrogram and the global physiological signal modulation curve, illustrating the underlying respiratory amplitude variations and cardiac frequency drifts throughout the free-breathing acquisition.

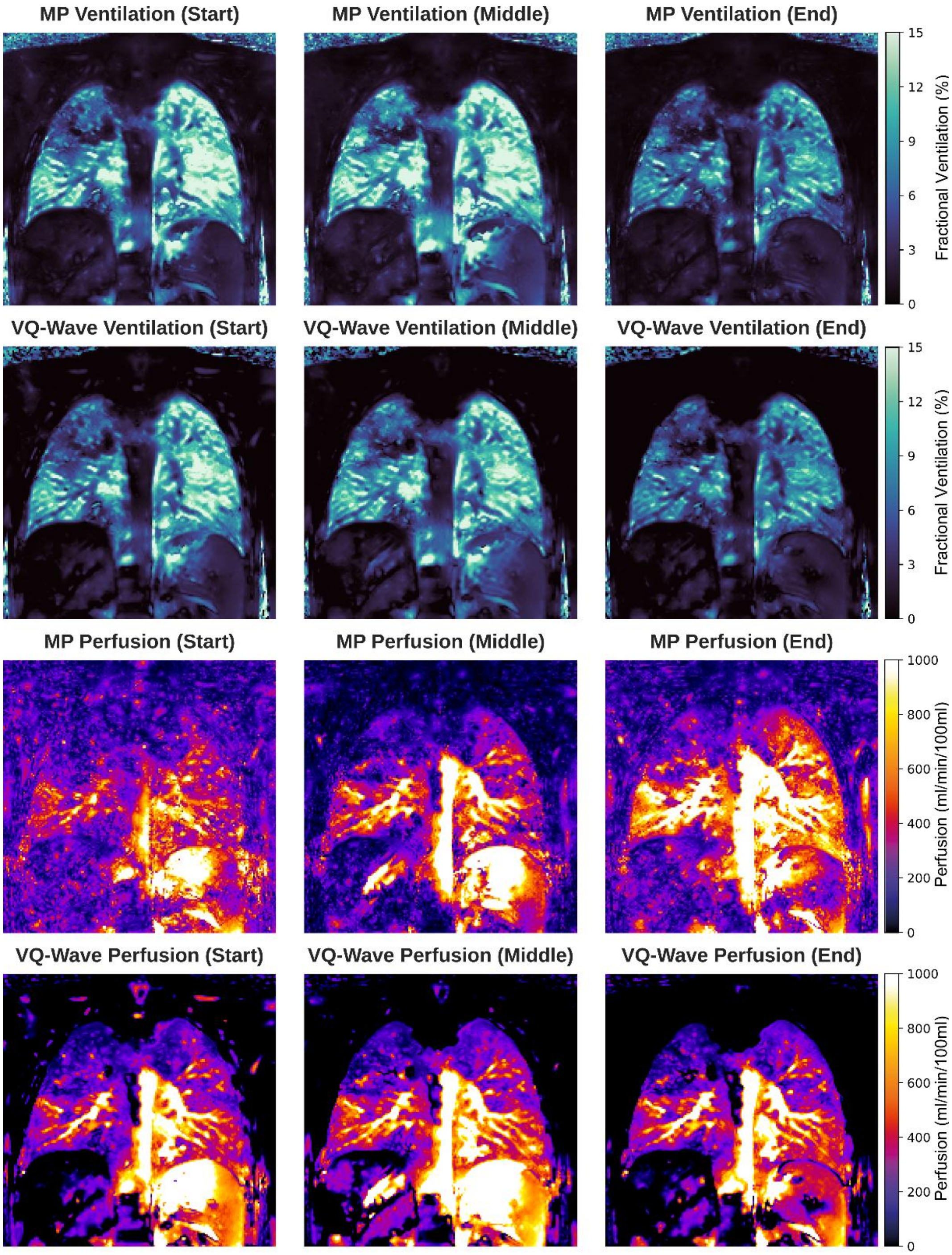

**Figure 9.** Temporal reproducibility of functional reconstructions from truncated acquisitions. Comparison of fractional ventilation and perfusion maps generated using three distinct 40-frame windows extracted from beginning, middle and end of the full-length free-breathing cystic fibrosis patient acquisition. While VQ-Wave maintains consistent structural and quantitative fidelity across all temporal windows, the matrix pencil (MP) reconstructions exhibit pronounced window-dependent instability and artifactual variations.

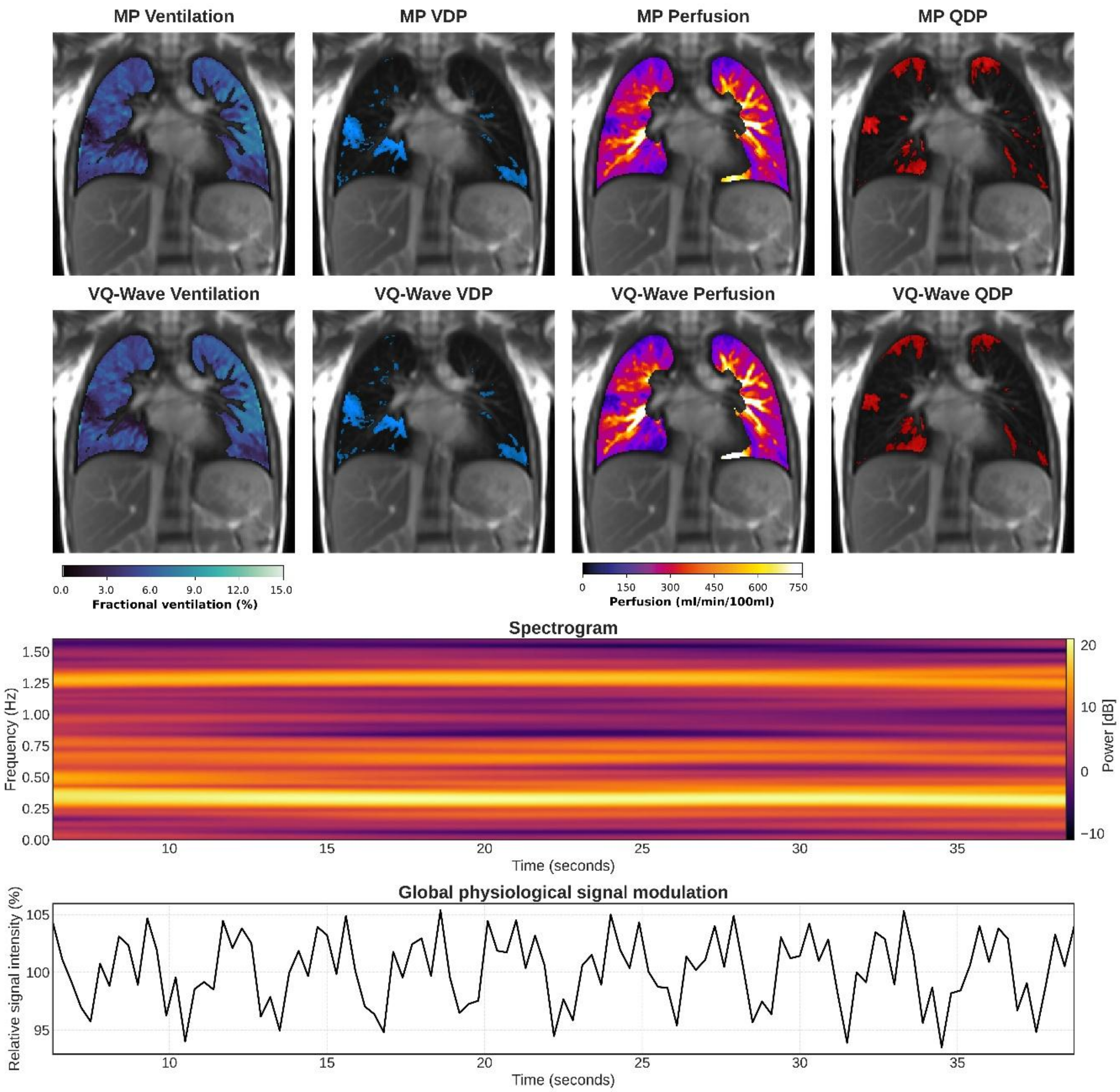

**Figure 10.** Diagnostic defect mapping under stable physiological conditions. Evaluation of a 12-year-old female with cystic fibrosis exhibiting stable cardiopulmonary dynamics. Clinical lung function tests indicated a forced vital capacity (FVC) of 82%, forced expiratory volume in 1 second (FEV1) of 82%, and a lung clearance index at 2.5% (LCI 2.5%) of 7.68. (Top rows) Comparison of fractional ventilation and perfusion maps alongside their corresponding defect percentage maps (VDP and QDP) generated by the matrix pencil (MP) reference and VQ-Wave. Under these optimal conditions, both methods successfully identify the exact same complex peripheral functional deficits, yielding highly comparable global defect burdens (VQ-Wave: VDP = 16.9%, QDP = 15.3%; MP: VDP = 16.5%, QDP = 17.9%). (Bottom rows) The time-frequency spectrogram and global physiological signal modulation curve confirm the presence of stable, periodic respiratory and cardiac cycles with minimal frequency drift or amplitude modulation.

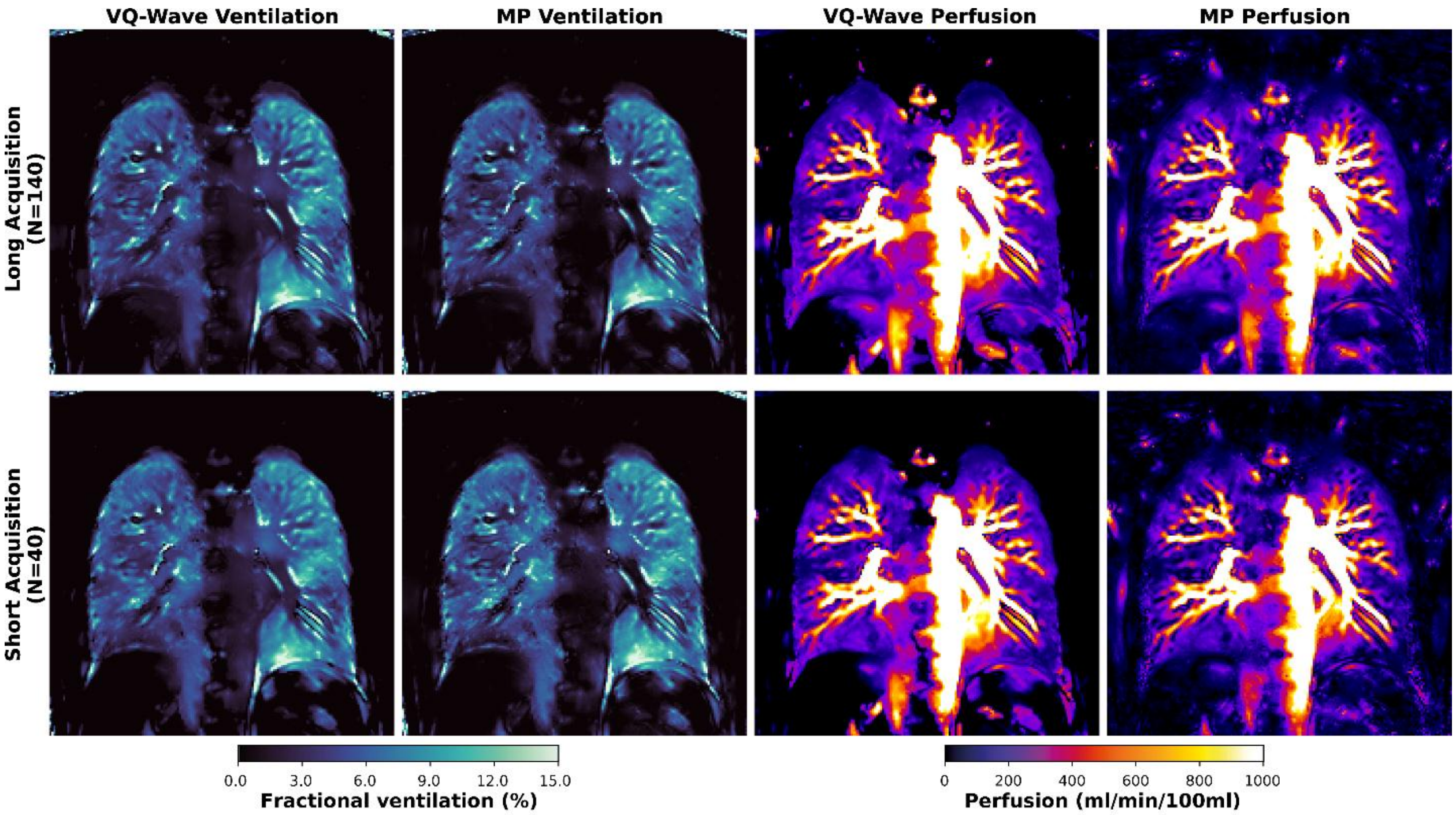

**Supporting Information Figure S1.** Visual comparison of reconstruction stability in a healthy volunteer. Comparison of fractional ventilation (left) and perfusion (right) maps reconstructed using a standard long acquisition (N=140, top row) versus an accelerated short acquisition (N=40, bottom row). VQ-Wave and the matrix pencil (MP) reference produce high-quality functional maps from the full time-series. When the acquisition is reduced to 15 seconds (N=40), VQ-Wave maintains diagnostic image quality with no loss of structural detail. In contrast, the MP perfusion map exhibits visible degradation, including signal dropout in the lower lobes and increased background noise, indicating a failure to resolve the cardiac signal from the limited data.

| Tissue Class / Feature | Signal characteristics | Parameter Ranges / Probabilities | Training goal |
|---|---|---|---|
| **Lung parenchyma** | Low to moderate baseline, variable $A_v$ and $A_q$ | **Probability:** 40%<br><br>**Baseline:** 15–250 a.u. | Learning simultaneous ventilation and perfusion separation |
| **Pulmonary vessels** | High baseline, strong $A_q$, $A_v \approx 0$. Includes “weak vessel” variants to improve sensitivity | **Probability:** 30%<br><br>**Baseline:** 200–800 a.u. | Learning that high signal intensity does not imply ventilation; identifying pure perfusion sources |
| **Static tissue** | High baseline, zero dynamic amplitude ($A_v = A_q = 0$) | **Probability:** 20%<br><br>**Baseline:** 300–800 a.u. | Teaches the network to predict zero-amplitude in static regions (e.g. muscle, fat) |
| **Air background** | Very low baseline signal and absence of any V/Q modulation ($A_v = A_q = 0$) | **Probability:** 10%<br><br>**Baseline:** 3–20 a.u. | Ensures robustness against low-SNR / random phase noise |
| **Physiological frequencies** | Shared underlying respiratory and cardiac rates applied across the entire field of view | **Ventilation ($f_v$):** 0.10–0.60 Hz<br><br>**Perfusion ($f_q$):** 0.75–2.0 Hz | Trains the network to track temporal dynamics independently of local tissue amplitudes |
| **Pathological defects** | Regional absence of $A_v$ or $A_q$ modulation within otherwise standard lung parenchyma | **Ventilation Defect:** 10% probability<br><br>**Perfusion Defect:** 10% probability | Force the network to correctly identify physiological voids (e.g., embolisms, air trapping) |
| **Irregular dynamics & noise** | Non-stationary baseline drift, erratic amplitude spikes, and noise | **Baseline Drift:** Additive Gaussian random walk.<br><br>**Sighs/Spikes:** up to 50% amplitude increase<br><br>**Amplitude Jitter:** Random heart beat-to-beat variance<br><br>**Frequency Drift:** Non-stationary variation in $f_v$ and $f_q$<br><br>**Noise ($\sigma$):** 1.0–17.5 | Ensure temporal filter stability against patient motion, deep sighs, heart rate variability, shifting respiratory rates or baseline drifts |

**Supporting Information Table S1.** Physiological parameters, signal characteristics, and training goals of the synthetic data generator**.** Note: The assigned probabilities reflect the sampling frequency of the center voxel’s tissue class during stochastic training. Parameter bounds, pathological defect rates, and dynamic noise profiles were explicitly defined to force the network to generalize across diverse physiological conditions and bridge the clinical domain gap.

| Category | Parameter | Value / Setting |
| --- | --- | --- |
| **Optimization** | Optimizer | AdamW |
| | Base learning rate | $1 \times 10^{-3}$ |
| | Weight decay | $1 \times 10^{-4}$ |
| | Number of samples | $10^6$ spatial patches |
| | Training schedule | 50 epochs |
| | Batch size | 1024 spatial patches |
| | Learning rate scheduler | Cosine annealing warm restarts ($T_0$=10, $T_{mult}$=2) |
| **Loss functions** | Amplitude targets $(A_v, A_q)$ | Smooth L1 loss |
| | Frequency & phase targets $(f_v, f_q, \phi_v, \phi_q)$ | Mean squared error (MSE) |
| **Acquisition bounds** | Temporal resolution | Uniformly sampled in [0.20, 0.33] seconds |
| | Sequence length | Uniformly sampled [40, 190] frames |
| **Spatial augmentation** | Local 3×3 patch rotation | Random 90° rotations ($k \in [0,1,2,3]$) |
| | Local 3×3 patch flipping | Horizontal and Vertical flips (p = 0.5) |

**Supporting Information Table S2.** Summary of network optimization and training hyperparameters. The table details the specific machine learning configurations, including the optimization schedule, loss function definitions, temporal acquisition boundaries, and the spatial augmentation strategies